\documentclass{iaus}
\usepackage{graphicx}

\title[Pulsar timing array projects] %% give here short title %%
{Pulsar timing array projects}

\author[G. Hobbs]   %% give here short author list %%
{G. Hobbs$^1$
\affiliation{$^1$ Australia Telescope National Facility, CSIRO, P.O. Box 76, Epping, NSW 1710, Australia} \\email: {\tt george.hobbs@csiro.au}}

\pubyear{2008}
\volume{xxx}  %% insert here IAU Symposium No.
\pagerange{119--126}
\jname{Title of your IAU Symposium}
\editors{A.C. Editor, B.D. Editor \& C.E. Editor, eds.}
\begin{document}

\maketitle

\begin{abstract}
Pulsars are amongst the most stable rotators known in the Universe.  Over many years some millisecond pulsars rival the stability of atomic clocks.  Comparing observations of many such stable pulsars may allow the first direct detection of gravitational waves, improve the Solar System planetary ephemeris and provide a means to study irregularities in terrestrial time scales.  Here we review the goals and status of current and future pulsar timing array projects.

\end{abstract}

\firstsection % if your document starts with a section,
\section{Introduction}

The pulsar timing technique (e.g. Lorimer \& Kramer 2005 for an overview and Edwards, Hobbs \& Manchester 2006 for a detailed description) has enabled some of the most exciting recent results in astrophysics.  For instance, the first extra-Solar planets were discovered orbiting the pulsar PSR~B1257$+$12 (Wolszczan \& Frail 1992), the first evidence for gravitational waves was provided by the binary system PSR~B1913$+$16 (Hulse \& Taylor 1974) and the most stringent tests of general relativity in the strong-field regime have been carried out using the double pulsar system (Kramer et al. 2006).   Even though the PSR~B1913$+$16 system has provided evidence for the existence of gravitational waves, most researchers regard this as indirect evidence and not a direct detection of such waves.

\begin{figure*}
\begin{center}
\includegraphics[width=5cm,angle=-90]{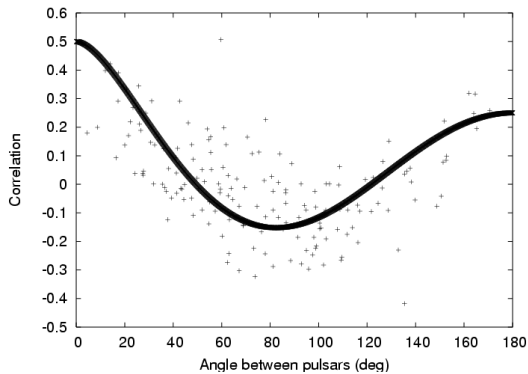}
\end{center}
\caption{The expected correlation in the timing residuals of pairs of pulsars as a function of angular separation for an isotropic gravitational wave background.  The solid line gives the theoretical curve.  The dots are calculated by correlating simulated data sets for 20 pulsars in the presence of a gravitational wave background that produces white timing residuals.}\label{fg:corr}
\end{figure*}

Sazhin (1978) and Detweiler (1979) showed that low-frequency gravitational waves will induce pulsar timing residuals\footnote{Pulse times of arrival for a given pulsar are compared with a physical model of the pulsar.  The differences between the measured arrival times and predicted arrival times using the model are known as the pulsar timing residuals and denote physical effects that are not included in the model.} that could be detectable with high-precision observations of the pulse arrival times.  However, with a single pulsar it is difficult, and perhaps impossible, to distinguish between the effects of gravitational waves and many other effects such as the irregular rotation of the neutron star, interstellar medium propagation effects or inaccuracies in the planetary ephemeris.  Hellings \& Downs (1983) showed that such effects could be distinguished by looking for correlated behaviour in the timing residuals of many pulsars.  In brief, the timing residuals caused by irregular rotation or propagation effects should be uncorrelated between different pulsars.  Irregularities in the terrestrial atomic time scale would produce completely correlated timing residuals.  The angular correlation function expected for an isotropic, stochastic gravitational wave background (Hellings \& Down 1983) is shown in Figure~\ref{fg:corr} (using simulated data described in Hobbs et al. 2009).

The first attempt to create a ``pulsar timing array'' in which enough pulsars are observed with sufficient timing precision to search for correlated signals was reported by Foster \& Backer (1990).  More recently Jenet et al. (2005) calculated the number of pulsars, the timing precision and data span required in order to make a significant detection of the expected stochastic background of gravitational waves.   In summary, at least 20 millisecond pulsars are required, observed for five years with timing precision around 100\,ns. In this review paper we discuss the current status of various existing pulsar timing array projects that aim to achieve this goal.  In the future, telescopes such as the Square Kilometre Array may make gravitational wave detection using pulsars commonplace.  The current status of this project is described in \S\ref{sec:future}.

% GW
% Aims of the projects

\section{The pulsar timing array projects}

The main aims of pulsar timing arrays are to search for correlated signals in the pulsar timing residuals in order to 1) improve the Solar System ephemeris, 2) look for irregularities in the terrestrial time standards and 3) make a direct detection of gravitational waves.  Three projects have been started:

\begin{itemize}
\item{\emph{The Parkes Pulsar Timing Array} is the only timing array project (to date) in the Southern Hemisphere.   This project makes use of the 64-m Parkes radio telescope.  Every two--three weeks, each of 20 pulsars is observed for approximately one hour using a dual-band 10cm-50cm receiver and for another hour using a 20cm receiver.  The data are processed using the PSRCHIVE suite of software (Hotan, van Straten \& Manchester 2004) that allows polarimetric and flux density calibration.  Pulsar timing models and the resulting timing residuals are obtained using \textsc{tempo2} (Hobbs, Edwards \& Manchester 2006).}

\item{\emph{The European Pulsar Timing Array} is making precise timing observations of approximately 15 pulsars.  This project makes use of the existing four 100-m class telescopes in Europe (Jodrell Bank, Effelsberg, Westerbork and Nancay) and will also include observations from the 64-m Sardinian radio telescope after its completion.  As well as using each of these telescopes as individual instruments, the Large European Array for Pulsars (LEAP) project will allow these telescopes to act as a phased array giving the equivalent sensitivity of a 200\,m telescope.}

\item{\emph{The North American Pulsar Timing Array}, known as \emph{Nanograv}, combines data collected with the 300-m Arecibo radio telescope and the 100\,m GreenBank telescope.  These telescopes are currently obtaining about monthly observations on 24 pulsars.} 

\end{itemize}

Members of these three projects have agreed to share data to form the International Pulsar Timing Array (IPTA) project which hopefully will result in the detection of gravitational waves within 5-10 years.

% Give limits versus time
% Have plot of timing precision versus time

\subsection{The planetary ephemeris}

 \begin{figure*}
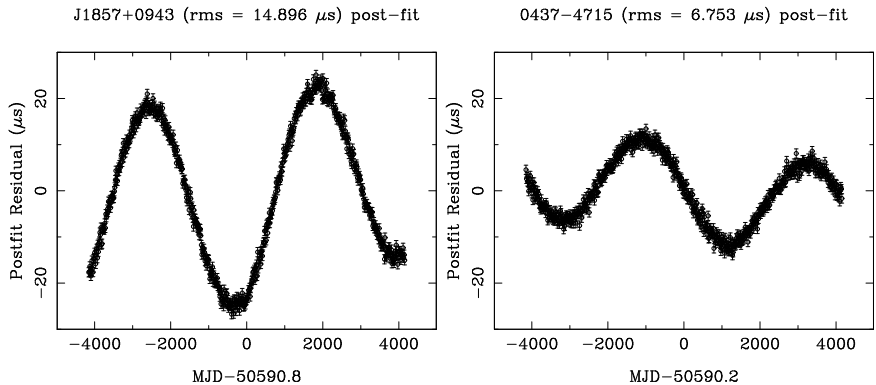

\begin{center}
\includegraphics[width=5cm,angle=-90]{plot1.ps}
\includegraphics[width=5cm,angle=-90]{plot2.ps}
\end{center}
\caption{Simulated timing residuals for PSR~J1857$+$0943 (left panel) and J0437$-$4715 induced by an error in the mass of Jupiter of $10^{-8}$\,M$_\odot$.}\label{fg:jupiter}
\end{figure*}

The pulsar timing procedures requires that the observed pulse times-of-arrival are corrected to the Solar System barycentre.  This correction is carried out using a planetary ephemeris that provides the position of the Earth with respect to the barycentre.  Such ephemerides are complex and take into account the masses and orbital parameters of the planets and over 300 asteroids.  If the mass of one planet was slightly incorrect, then the position of the barycentre will be incorrectly determined and sinusoidal timing residuals will be induced with the period of the planet's orbit (note that a change in the mass of Jupiter will  affect the entire fit that is carried out to create the planetary ephemeris.  The change in position of the barycentre is a first order effect and, as the change in Jupiter's mass considered here is small, is the only change in the ephemeris that is considered.  A more detailed study will be published elsewhere.)   In Figure~\ref{fg:jupiter} we simulate the timing residuals of PSR~J1857$+$0943 (left panel) and PSR~J0437$-$4715 (right panel) for a change in the mass of Jupiter of 10$^{-8}$\,M$_\odot$.  The IPTA project is currently combining data from a selection of pulsars to place a limit on the error in Jupiter's mass which is expected to be more precise than measurements obtained using the \emph{Voyager} space-craft, but not as precise as unpublished \emph{Galileo} space-craft measurements.  In the near future pulsar timing arrays should be able to rule out, or detect, the existence of an Earth-mass ``planet-X'' closer than 60\,A.U., or a Jupiter-mass object closer than 300\,A.U.

\subsection{Terrestrial time standards}

\begin{figure*}
\begin{center}
\includegraphics[width=5cm,angle=-90]{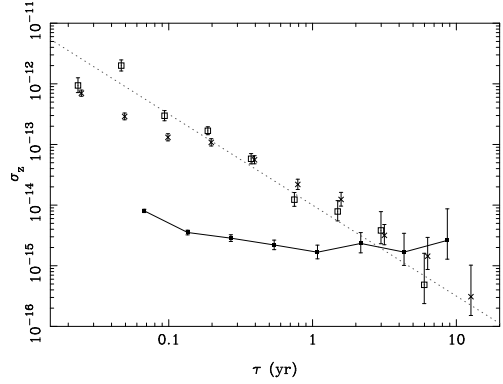}
\end{center}
\caption{$\sigma_z(\tau)$ for PSR~J0437$-$4715 and PSR~J1909$-$3744 and for a data sequence of atomic clock differences: PTB-NIST.}\label{fg:clock}
\end{figure*}

Millisecond pulsars are extremely stable. A measure of stability, $\sigma_z$ (Matsakis et al. 1997), allows pulsars to be compared with terrestrial time scales.  In Figure~\ref{fg:clock} we plot the stability of PSRs~J0437$-$4715 (crosses) and J1909$-$3744 (squares).  The arrival times for both these pulsars were referenced to the Bureau International des Poids et Mesures (BIPM) 2008 terrestrial time standard.   The solid line indicates the stability of the difference between the Physikalisch-Technische Bundesanstalt (PTB) and National Institute of Standards and Technology (NIST) time standards. The dotted line corresponds to timing residuals with a flat spectrum.   It is clear that on short time scales the atomic time scales are significantly more stable than pulsars, but on time scales of $\sim 10$\,yr and longer the pulsars become as stable as the terrestrial clocks.  In the near future it is expected that correlated signals in the timing residuals will indicate the presence of irregularities in the terrestrial time scales and will allow the formation of a pulsar-based time standard (e.g. Rodin 2008).
 
\subsection{Gravitational wave detection}

\begin{figure*}
\begin{center}
\includegraphics[width=5cm,angle=-90]{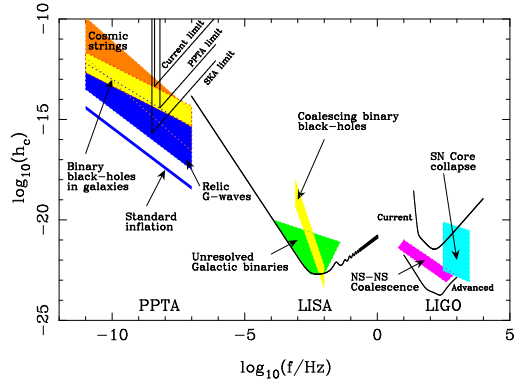}
\end{center}
\caption{Characteristic strain sensitivity for existing and proposed GW detectors as 
a function of GW frequency. Predicted signal levels from various astrophysical sources 
are shown (Hobbs 2008).}\label{fg:sens}
\end{figure*}

The most likely detectable gravitational wave signal is from an isotropic, unpolarised, gravitational-wave background caused by binary supermassive black holes at the centre of merged galaxies (Sesana et al. 2008).  The induced timing residuals induced by such a background can be simulated using the TEMPO2 software package (Hobbs et al. 2009) and lead to upper bounds on the amplitude of any such signal.  Jenet et al. (2006) gave an upper limit on the energy density per unit logarithmic frequency interval of $\Omega_g[1/(8{\rm yr})]h^2 \leq 1.9\times 10^{-8}$ which can be used to constrain the merger rate of supermassive binary black hole systems at high redshift (Wen et al. 2009).  A comparison between the sensitivity to gravitational waves for different detectors is shown in Figure~\ref{fg:sens} along with the predicted signal levels from various astrophysical sources.

\section{Future timing array projects}\label{sec:future}

With current telescopes it is a challenge to achieve a timing precision of $\sim$100\,ns for more than a few pulsars (the best published data set is for PSR~J0437$-$4715 observed for more than 10\,yr with rms timing residuals $\sim 200$\,ns; Verbiest et al. 2008).  Within $\sim$10\,yr it is expected that various new telescopes such as the Australian Square Kilometre Array Pathfinder (ASKAP) and the South African pathfinder (MEERKAT) telescopes will increase the number of known pulsars and provide a larger number of telescopes in the Southern Hemisphere for high-precision pulsar timing (Johnston et al. 2008, http://www.ska.ac.za). Also a new 500m-diameter telescope (FAST) should be completed in China (Nan et al. 2006).  This telescope will be a highly efficient pulsar search and timing instrument and should significantly improve our timing precision for a large number of pulsars.  

On a longer timescale, the Square Kilometre Array (SKA) telescope is planned.  This telescope should be able to observe many hundreds of pulsars with the precision currently achieved for only a few pulsars.  Such data sets should allow the gravitational-wave background and individual sources of such waves to be studied in detail (Cordes et al. 2004, Kramer et al. 2004).  For instance a high S/N detection of a stochastic background would provide a test of general relativity (Lee et al. 2008) and allow a detailed understanding of the properties of the sources that form the background.

\section{Conclusion}

Many pulsars are currently being observed as part of pulsar timing array projects.  Correlated timing residuals allow 1) irregularities in terrestrial time standards to be identified, 2) inaccurate planetary masses in the Solar System ephemeris to be corrected and 3) the detection of gravitational wave signals.  Combining observations from Northern and Southern hemisphere telescopes should allow the detection of gravitational waves on a time scale of 5-10 years.

\section{Acknowledgements}

GH is supported by an Australian Research Council QEII Fellowship (project \#DP0878388).


\begin{thebibliography}{}

\bibitem[]{cor04}{Cordes, J.,  Kramer, M., Lazio, T. J. W., Stappers, B. W., Backer, D. C. \& Johnston, S.} 2004, New Astronomy Reviews, 48, 1413

\bibitem[]{det79}{Detweiler, S.} 1979,
\textit{ApJ}, 234, 1100

\bibitem[]{ehm06}{Edwards, R., Hobbs, G. \& Manchester, R.} 2006,
\textit{MNRAS}, 372, 1549

\bibitem[]{fb90}{Foster, R. S. \& Backer, D. C.} 1990,
\textit{ApJ}, 361, 300

\bibitem[]{hd83}{Hellings, R. W. \& Downs, G. S.} 1983,
\textit{ApJ}, 265, L39

\bibitem[]{hem06}{Hobbs, G., Edwards, R. \& Manchester, R.} 2006,
\textit{MNRAS}, 369, 655

\bibitem[]{hob08}{Hobbs, G.} 2008,
\textit{Classical and Quantum Gravity}, 25, 11

\bibitem[]{hob09}{Hobbs, G. et al.} 2009,
\textit{MNRAS}, 394, 1945

\bibitem[]{hvm04}{Hotan, A. W., van Straten, W. \& Manchester, R. N.} 2004,
\textit{PASA}, 21, 302

\bibitem[]{ht74}{Hulse, R. A. \& Taylor, J. H.} 1974,
\textit{ApJ}, 191, L59

\bibitem[]{jhlm05}{Jenet, F. A., Hobbs, G., Lee, K. J. \& Manchester, R. N.} 2005,
\textit{ApJ}, 625, L123

\bibitem[]{jen06}{Jenet, F. A. et al.} 2006,
\textit{ApJ}, 653, 1571

\bibitem[]{jtb+08}{Johnston, S.. et al.} 2008,
\textit{ExA}, 22, 151

\bibitem[]{kbc+04}{Kramer, M., B. W., Backer, D. C., Cordes, J.,  Lazio, T. J. W., Stappers  \& Johnston, S.} 2004, New Astronomy Reviews, 48, 993

\bibitem[]{ksm+06}{Kramer, M. et al.} 2006,
\textit{Sci}, 314, 97

\bibitem[]{ljp08}{Lee, K. J., Jenet, F. A. \& Price, R. H.} 2008,
\textit{ApJ}, 685, 1304

\bibitem[]{lk05}{Lorimer, D. R. \& Kramer, M.} 2005,
\textit{Handbook of Pulsar Astronomy}, Cambridge University Press

\bibitem[]{mte97}{Matsakis, D. N., Taylor, J. H, \& Eubanks, T. M.} 1997,
\textit{A\&A}, 326, 924

\bibitem[]{nwz+06}{Nan, R. D., Wang, Q. M., Zhu, L. C., Zhu, W. B., Jin, C. J. \& Gan, H. Q.} 2006,
\textit{Chin. J. Astron. Astrophys., Suppl.}, 6, 304

\bibitem[]{rod08}{Rodin, A.} 2008,
\textit{MNRAS}, 387, 1583

\bibitem[]{saz78}{Sazhin, M. V.} 1978,
\textit{SvA}, 22, 36

\bibitem[]{svc08}{Sesana, A., Haardt, F. \& Madau, P.} 2008,
\textit{MNRAS}, 390, 192

\bibitem[]{vbs+08}{Verbiest, J. et al.} 2008,
\textit{MNRAS}, 679, 675

\bibitem[]{wen09}{Wen, Z. et al.} 2009, \emph{``Constraining the coalescence rate of supermassive black-hole binaries using pulsar timing''}, submitted to MNRAS

\bibitem[]{wf92}{Wolszczan, A. \& Frail, D. A.} 1992,
\textit{Nature}, 355, 145







\end{thebibliography}
\end{document}